\documentclass[12pt,fullpage]{article}
\usepackage {epsfig}
\usepackage {bbbold}
\usepackage {amssymb}
\def\slantfrac#1#2{\hbox{$\,^{#1}\!/_{#2}$}}
\def\mdot{{\raisebox{1pt}{\hbox{$\stackrel{\bullet}{M}$}}}\ \!\!}
\def\apgt{\ {\raise-.5ex\hbox{$\buildrel>\over{\scriptstyle\sim}$}}\ }
\def\aplt{\ {\raise-.5ex\hbox{$\buildrel<\over{\scriptstyle\sim}$}}\ }

\begin{document}

\title{3D Gasdynamic Modelling of the Changes in the Flow Structure
During Transition From Quiescent to Active State in Symbiotic Stars}

\author{M. Mitsumoto$^1$, B. Jahanara$^1$, T. Matsuda$^1$,\\
K. Oka$^2$, D.V. Bisikalo$^3$, E.Yu. Kilpio$^3$, H.M.J. Boffin$^4$,\\
 A.A. Boyarchuk$^3$, O.A. Kuznetsov$^{3,5}$\\[5mm]
$^1$ Department of Earth and Planetary Sciences,\\
Kobe University,Kobe 657-8501, Japan\\
$^2$ Mizuho Information and Research Institute, Inc.,\\
Tokyo 101-8443, Japan\\
$^3$ Institute of Astronomy RAS, Moscow, Russia\\
$^4$ European Southern Observatory, Karl-Schwarzschild-Str. 2,\\
D-85738 Garching, Germany\\
$^5$ Keldysh Institute for Applied Mathematics, Moscow, Russia\\}

\date{}

\maketitle

\begin{abstract}

The results of 3D modelling of the flow structure in the classical
symbiotic system Z~Andromedae are presented. Outbursts in systems
of this type occur when the accretion rate exceeds the upper limit
of the steady burning range. Therefore, in order to realize the
transition from a quiescent to an active state it is necessary to
find a mechanism able to sufficiently increase the accretion rate on a time scale
typical to the duration of outburst development.

Our calculations have confirmed the transition mechanism from
quiescence to outburst in classic symbiotic systems suggested
earlier on the basis of 2D calculations (Bisikalo et al, 2002).
The analysis of our results have shown that for wind velocity of
20~km/s an accretion disc forms in the system. The accretion rate
for the solution with the disc is $\sim22.5-25\%$ of the mass loss
rate of the donor, that is,~ $\sim4.5-5\cdot10^{-8}$ $M_\odot$/yr
for Z~And. This value is in agreement with the steady burning
range for white dwarf masses typically accepted for this system.
When the wind velocity increases from 20 to 30 km/s the accretion
disc is destroyed and the matter of the disc falls onto the
accretor's surface. This process is followed by an approximately
twofold accretion rate jump. The resulting accretion rate growth
is sufficient for passing the upper limit of the steady burning
range, thereby bringing the system into an active state. The time
during which the accretion rate is above the steady burning value
is in a very good agreement with observations.

The analysis of the results presented here allows us to conclude that
small variations in the donor's wind velocity can lead to the
transition from the disc accretion to the wind accretion and, as a
consequence, to the transition from quiescent to active state in
classic symbiotic stars.

\end{abstract}

\section{Introduction}

Symbiotic stars are characterized by peculiar spectra where
molecular absorption bands - the characteristic features of a
cool giant - are present together with emission lines corresponding
to a high excitation level. The red and infrared spectra of symbiotic
stars are typical of cool giants while in the UV range they are
characterized by a very hot continuum. It is widely assumed
that most symbiotic stars are detached binaries
consisting of a cool giant and a white dwarf surrounded by a
nebulosity [\ref{Boyar67}]. Mass exchange in these systems is
driven by stellar wind.

The goal of this work is to carry out a study of classical symbiotic
systems using as example one of the most well-studied
representatives of this class - Z~And. According to the analyses of
its energy distribution in a wide spectral range [\ref{FC88}],
the components of Z~And have the following characteristics: the
cool M3.5III giant has a mass $\sim2~M_\odot$ and a radius
$\sim$ 100$R_\odot$, while the white dwarf has a mass
$\sim0.6~M_\odot$, a radius $\sim$ 0.07$R_\odot$ and a
temperature $\sim 10^5$~K. The giant loses mass at the rate of
$\sim2\times10^{-7}~M_\odot$/yr. Gas in the circumbinary envelope
has an electron density of $\sim2\cdot10^{10}~\mbox{cm}^{-3}$ and
a temperature $\sim1.5\div8\cdot10^4$~K. The separation of the
system is 482$R_\odot$, the orbital period is 758 days. More
than 100 years of observations have shown that this star presents outbursts.
The last of them took place in 2000 and has
been actively observed at different wavelengths.

Following the characteristics of their outbursts, symbiotic stars
can be divided in two types [\ref{PR80}, \ref{MK92}]. The first
type includes so-called classical symbiotics (e.g. Z And, AG Peg,
AG Dra, CI Cyg, AX Per). Their outbursts typically last a few
months and have amplitudes of 2-4$^{\rm m}$.
The typical feature of such outbursts is the weakening of high
excitation lines while the brightness increases. The detailed
description of the behaviour of these stars is given in
[\ref{Boyar69}, \ref{Kenyon86}]. Symbiotic novae (e.g. V1016 Cyg,
V1329 Cyg, RS Oph, HM Sge) show the second type of outbursts.
These outbursts are longer and brighter. Their energetics can
even exceed that of novae outbursts. The principal difference
of these outbursts from those of the first type is the increase of
the ionization degree while the brightness increases. The behaviour of
these stars during outbursts is described in [\ref{Baratta},
\ref{Mammano}, \ref{Ciatti},\ref{Puetter}]. The second type of
outbursts is similar to slow classical novae outbursts
[\ref{Allen}, \ref{Kenyon83}] (for differences between symbiotic
novae and slow classical novae outbursts see [\ref{IbenTu96}]).

The most probable mechanism describing the observational
manifestations of classical symbiotics as well as of symbiotic
novae is thermonuclear burning on the accretor's surface [see,
e.g. \ref{Kenyon86}, \ref{IbenTu96}]. It has been shown in various
works [\ref{TU}, \ref{Zytkow}, \ref{PR80}, \ref{Iben},
\ref{IbenTu89}] that the process of thermonuclear burning of
hydrogen on the white dwarf surface strongly depends on the
accretion rate $\mdot^{accr}$ [\ref{PR80}]. Only, in a narrow
range of $\mdot^{accr}$ is the steady hydrogen burning possible
[\ref{Zytkow}, \ref{Sion}, \ref{Fuji}]. The lower limit of this
range is given by the expression [\ref{Sienk},\ref{Iben}]

$$
\mdot_{steady,min}=1.3\cdot10^{-7}\cdot
\left(\frac{M}{M_\odot}\right)^{3.57}M_\odot/\mbox{yr}\,,
$$
while the higher limit is given by [\ref{Pacz}, \ref{Uus}]

$$
\mdot_{steady,max}=6\cdot10^{-7}\cdot
\left(\frac{M}{M_\odot}-0.522\right)M_\odot/\mbox{yr}\,,
$$
For Z~And where $M=0.6M_\odot$, steady burning is possible in the
range $2.1\cdot10^{-8}M_\odot/\mbox{yr}
\aplt\mdot^{accr}\aplt4.7\cdot10^{-8}M_\odot/\mbox{yr}$. If the
accretion rate is below this interval, thermonuclear burning
exhausts the matter faster than accretion can replenish it
bringing the burning to a halt. Following this, hydrogen begins to
accumulate on the white dwarf surface until the pressure near the
hydrogen envelope base reaches a critical value and a  hydrogen
shell flash occurs. The bolometric luminosity increases by a
factor 10--100 times in the course of $\sim$ 1 year and stays at
the active state for $\sim$ 10 years. The total duration of the
outburst is dozens of years that corresponds to the nuclear time
scale. This is observed in symbiotic novae.

If for some reason $\mdot^{accr}$ becomes larger than
$\mdot_{steady,max}$, the accreted matter accumulates above the
burning shell and expands to giant dimensions. Such behaviour is
typical for classical symbiotic stars. Though the bolometric
luminosity stays constant, the visual brightness increases by 1-2
magnitudes and the effective temperature decreases. Optical
outburst develops on the thermal timescale. In accordance with
existing models of thermonuclear burning on a white dwarf' surface
[\ref{Sienk},\ref{Zytkow}], small variations of the accretion rate
can lead to significant changes of the temperature at
practically constant bolometric luminosity.
For a white dwarf of $1.2M_\odot$ mass, an
envelope mass change of $4\times10^{-7}~M_\odot$ can result in
a 3$^{\rm m}$ luminosity increase [\ref{PR80}]. This value corresponds to the
typical amplitude of Z~And outbursts.

Z~And belongs to the class of classical symbiotics and, in
accordance with the accepted model, the accretion rate in the
quiescent state should be in the steady burning range. For an
outburst to develop the accretion rate should become large enough
to reach a value above the steady burning range. It should be
    noted that the time scale of the accretion rate increase should
    correspond to the characteristic time of outburst development
($\sim100$ days). Such behaviour of the accretion rate in a binary
system is possible in the framework of the mechanism proposed in
[\ref{Bisikalo2002}] using results of 2D gasdynamic modelling. The
main idea of the mechanism is that even minor variations in the
donor's wind velocity is enough to lead to the accretion regime
change - from disc accretion to wind accretion. During the
transition period, namely during the disc destruction the
accretion rate increases abruptly and exceeds the upper limit of
the steady burning range. The goal of this work is to check the
possibility of realizing such a mechanism using a more realistic
3D model.

\section{The model}

In  order  to  study  the  gas  flow  structure  in  the symbiotic
system Z And 3D numerical simulations have been carried out. The
zero point of the coordinate system was placed at the centre of
the accretor, the $x$- axis was directed along the line connecting
centres of the components and oppositely to the mass-losing star,
the $y$-axis  -- along the accretor orbital motion, and the
$z$-axis -- along the axis of rotation of the binary system. The
flow was described by the system of Euler equations in corotating
coordinate frame:

\[
{\frac{{\partial \rho} }{{\partial t}}} + {\frac{{\partial \rho
u}}{{\partial x}}} + {\frac{{\partial \rho v}}{{\partial y}}}+
{\frac{{\partial \rho w}}{{\partial z}}}=0\,,
\]

\[
{\frac{{\partial \rho u}}{{\partial t}}} + {\frac{{\partial (\rho
u^{2} + P)}}{{\partial x}}} + {\frac{{\partial \rho uv}}{{\partial
y}}} + {\frac{{\partial \rho uw}}{{\partial z}}} = - \rho
{\frac{{\partial\Phi} }{{\partial x}}} + 2\Omega v\rho\,,
\]

\[
{\frac{{\partial \rho v}}{{\partial t}}} + {\frac{{\partial \rho
uv}}{{\partial x}}} + {\frac{{\partial (\rho v^{2} +
P)}}{{\partial y}}} + {\frac{{\partial \rho vw}}{{\partial z}}} =
- \rho {\frac{{\partial\Phi} }{{\partial y}}} - 2\Omega u\rho\,,
\]

\[
{\frac{{\partial \rho w}}{{\partial t}}} + {\frac{{\partial \rho
uw}}{{\partial x}}} + {\frac{{\partial \rho vw}}{{\partial y}}} +
{\frac{{\partial (\rho w^{2} + P)}}{{\partial z}}} = - \rho
{\frac{{\partial\Phi} }{{\partial z}}}\,,
\]

\[
{\frac{{\partial \rho E}}{{\partial t}}} + {\frac{{\partial \rho
uh}}{{\partial x}}} + {\frac{{\partial \rho vh}}{{\partial y}}} +
{\frac{{\partial \rho wh}}{{\partial z}}} = - \rho
u{\frac{{\partial \Phi} }{{\partial x}}} - \rho v{\frac{{\partial
\Phi }}{{\partial y}}} - \rho w{\frac{{\partial \Phi }}{{\partial
z}}}\,.
\]
Here ${\bmath u}=(u,v,w)$ is the velocity vector, $P$ -- the pressure,
$\rho$ -- the density, $h=\varepsilon+P/\rho+|{\bmath u}|^{2}/2$ --
the specific  total enthalpy, $E=\varepsilon+|{\bmath u}|^{2}/2$ --
the specific  total energy, $\varepsilon$ -- the specific internal
energy, $\Omega $ -- the angular velocity of binary system's rotation,
and $\Phi({\bmath r})$ -- the force potential.

In the standard definition when only the gravitational forces from the
point mass components and the centrifugal force are taken into account,
the force potential is given by:

\[
\Phi({\bmath r})=-\frac{GM_1}{|{\bmath r}-{\bmath r}_1|}
-\frac{GM_2}{|{\bmath r}-{\bmath r}_2|}
-\slantfrac{1}{2}\Omega^{2}({\bmath r}-{\bmath r}_c)^2\,.
\]
Here $M_1$ is the mass  of  the accretor, $M_2$ -- the mass-losing
star's mass, ${\bmath r}_1$, ${\bmath r}_2$ -- the
radius-vectors of the centres of components, ${\bmath r}_c$ --
the radius-vector of the centre of mass of the system. This is
the so-called Roche potential. But  in  our case the additional force
responsible for the donor's wind acceleration should also be taken
into account.  Therefore, the form of the potential changes.
Previous studies (e.g.,
[\ref{GS}--\ref{Dima96}]) have shown that the general flow
structure in the system where components do not fill their Roche
lobes is defined first and foremost by the stellar wind
parameters. Since the mechanism of gas acceleration is poorly known
for cool giants, we mimic the radiation pressure by reducing the
gravitational attraction force to the donor star by a factor
($1-\Gamma$).
So, the modified force potential looks the following:

\[
\Phi({\bmath r})= -(1-\Gamma) \frac{GM_2}{|{\bmath r}-{\bmath
r}_2|} -\frac{GM_1}{|{\bmath r}-{\bmath r}_1|}
-\slantfrac{1}{2}\Omega^{2}({\bmath r}-{\bmath r}_c)^2\,.
\]

To close the system the perfect gas equation of state was used.

\[
P = (\gamma-1)\rho \varepsilon\,,
\]
The value of adiabatic index has been accepted $\gamma = 1.01$,
that corresponds to the case close to the isothermal one
[\ref{gamma1}--\ref{gamma3}].

We adopted for the parameters of the binary system those of Z And.
In our numerical simulations all variables were put in
dimensionless form as follows: the length was normalized by the
separation of two stars, $A$, and the time by $\Omega^{-1}$, and
so the velocity was normalized by $A\Omega$. The density was
normalized by that at the surface of the mass-losing star.

The method of computational fluid dynamics is described in
[\ref{Nagae}] and details are explained in
[\ref{Makita}, \ref{Fujiwara}]. We use the simplified flux vector
splitting (SFS) finite volume scheme to discretize the Euler
equation (the description of the SFS scheme is given in the
appendix of [\ref{Makita}]).

We assume a symmetry about the orbital plane, and therefore only compute
the upper half-space of the computational domain,
which is $-2<x<1$, $-1.5<y<1.5$, $0<z<1.0$. The region is divided into $%
253\times 253\times 85$ cells. The inner numerical boundary
surrounding the mass-losing star is represented by an
equipotential surface with a mean radius $R_{2n}$ (we set the
numerical boundary at a little larger radius, $R_{2n}=101R_{\odot}$,
rather than $R_{2}$). The radius of the mass-accreting star is too
small to fit in our numerical grid so it is represented by one
cell.

The cells just inside of the inner numerical boundary about the
mass-losing star are assumed to be filled with gas with
dimensionless density $\rho_{2}=1$, dimensionless sound speed
$c_{s}=0.178$ (corresponding to $ T_{2}=3200K$), and a
dimensionless normal speed to the surface, $V$. The real normal
velocity of the gas on the numerical boundary is determined by
solving Riemann problems between the cells adjacent to the inner
boundary surface. The cell representing the mass-accreting star
and the cells just outside of the outer numerical boundary are
filled by a gas with density $\rho _{0}=10^{-9}$, pressure
$p_{0}=10^{-8}/\gamma $, and three velocity components
$u_{0}=v_{0}=w_{0}=0$. This assumption does not mean that
accretion onto the mass accreting star nor that escape from the
computational domain do not occur. The velocity of the gas on the
boundaries are computed by solving Riemann problems. At the
initial stage $t=0$, the computational domain except the region
inside of the inner boundaries is assumed to be filled by a gas
with $\rho _{0}=10^{-9}$, $p_{0}=10^{-8}/\gamma $, and
$u_{0}=v_{0}=w_{0}=0.$ This gas is gradually replaced by the gas
supplied from the mass-losing star. We compute up to a few
rotation periods that is generally long enough for the system to
reach a quasi-steady state.

\section{Results}

The velocity of the mass-losing star's wind in
Z~And is $\sim$25 km/s [\ref{FC88}]. It
is presumed  in the framework of the considered mechanism that for
slightly smaller velocities an accretion disc will form
while for slightly greater values the disc will be
destroyed and a wind accretion type of flow will occur. In order to
check the applicability of the proposed mechanism, numerical
simulations for velocities of 25$\pm5$km/s have been carried
out.\footnote{Small variations of the wind velocity within $\pm$5
km/s can be easily explained by an activity of the giant
[\ref{MK92}].} Calculations for the wind velocity $V_w=20$ km/s
were necessary for studying the possibility of disc formation in
the system, accretion rate estimation and to check if the
accretion rate is indeed within the limits of the steady burning range.
Calculations for $V_w=30$ km/s were conducted in order to obtain
parameters of the flow structure in the system without disc. To
answer the questions concerning processes taking place after the
wind velocity change -- if the disc will be destroyed, how the
accretion rate will grow, if the accretion rate change will
go above the upper limit of the steady burning range --
calculations with a wind velocity jump from $20\rightarrow30$ km/s have
been carried out

\marginpar{\it\small\fbox{Fig.1}}

\marginpar{\it\small\fbox{Fig.2}}

The results of our numerical modelling have shown that for wind
velocity $V_w=20$ km/s and values of the parameter
$\Gamma\in[0.85;0.95]$ a steady accretion disc forms in the
system. In Figures 1,2 the flow structure for the case $V_w=20$
km/s and $\Gamma$=0.94 is shown. Density contours and velocity
vectors in the equatorial plane of the system are presented for
the whole computational domain (Fig. 1) and for the near-accretor
area $[-0.5A\ldots A]\times[-A\ldots 0.5A]$ (Fig. 2). The results
of the calculations are presented at the time  $t \sim 5 P_{orb}$
when a steady regime was already established in the system. These
results show that for this wind velocity value a bow shock and an
accretion disc of approximately $50-60 R_\odot$ form in the
system. Note, that two spiral density waves are seen in the
accretion disc. In this solution the accretion rate is
$\sim22.5-25\%$ of matter lost by the donor. Similar values of the
accretion efficiency for solution with accretion disc were
obtained by Mastrodemos and Morris [\ref{MM98}] for detached
binaries. For Z~And where the mass loss rate is estimated to be
$\sim2\cdot10^{-7}$ $M_\odot$/yr, the accretion rate will be
$\sim4.5-5\cdot10^{-8}$ $M_\odot$/yr, i.e. corresponding to the
steady burning range (near the upper limit).

 As one can see, due to the influence of the pressure gradient and
the presence of spiral shocks, the disc structure is far from
a conventional Keplerian disc. Since we used the full set of Euler
equations that incorporate advection term in energy equation, the
obtained flow structure is more similar to advection-dominated discs as
discussed by Walder and Folini [\ref{WF99}]. Of course, the
system of Euler equations with adiabatic energy equation does not
incorporate viscous heating nor radiative cooling,  but the
numerical viscosity, on the one hand, and the choice of the
adiabatic index $\gamma$ close to 1, on the other hand,
provide an implicit account of these processes. Thus, we believe
that our mathematical model is adequate to the physics of accretion
discs.

\marginpar{\it\small\fbox{Fig.3}}

\marginpar{\it\small\fbox{Fig.4}}

The solution with wind velocity $V_w=30$ km/s is presented in
Figures 3 and 4. Just as in Figures~1,2, here density contours and
velocity vectors are presented for all the computational domain
(Fig.~3) and for the area near the accretor $[-0.5A\ldots
A]\times[-A\ldots 0.5A]$ (Fig.~4). The situation presented
corresponds to the moment when the steady regime has already
become established. The analysis of these results show that in the
case when wind velocity equals 30 km/s the cone shock is close to
the accretor and does not leave any room for the formation of a
disc. In this solution a wind accretion type of flow instead of
disc accretion takes place with the accretion rate being
$\sim11-13\%$ of the matter leaving donor's surface. Similar flow
structures were obtained in 3D simulations performed by Dumm et
al. [\ref{Dumm}]. According to their results, the accretion
efficiency equals 6\% for the case of wind accretion.

\marginpar{\it\small\fbox{Fig.5}}

The analysis of results presented above allows to conclude that
even small change in the donor's wind velocity (within $\pm5$ km/s
around the observed value 25 km/s) leads to a flow structure and
accretion regime changeover, namely, to the transition from a disc
to a wind accretion flow. In order to consider the process of
transition between these two regimes after the wind velocity
increase, we took the stationary solution for the case 20 km/s
after $\sim 5 P_{orb}$ from the beginning of calculations
(situation presented in Figures 1a,1b) and then raised the
velocity up to 30~km/s. After the increase of the wind velocity on
the inner boundary the flow structure changes, evolving from a
state with an accretion disc to the one with a cone shock
presented in Figures 2a, 2b. Let us consider the process of the
flow restructuring. After $0.17P_{orb}$ ($\sim$133 days) since the
wind velocity change, the matter moving with increased velocity
reaches the vicinity of the accretor, namely, the bow shock
located in front of it. Then the wind continues to move further
crushing the accretion disc and making the matter of the disc fall
on the accretor's surface. A snapshot of this flow rearrangement
is presented in Figure 5. The area size and all the designations
are the same as in Figures 2,4. This situation corresponds to
$\sim$180 days after the wind velocity increase. It should be
mentioned that the study of the transition period is limited in
the framework of this model. After the accretion rate jump the
accepted boundary conditions on the accretor change and the model
used doesn't describe the real physical situation any more.
Correspondingly, the presented results of calculations of the flow
rearrangement period are correct only at first stages.

\marginpar{\it\small\fbox{Fig.6}}

\marginpar{\it\small\fbox{Fig.7}}

The behaviour of the accretion rate with time is shown in
Figures~4a,b. Here the time passed from the beginning of the
calculation in units of $t=P_{orb}/2\pi$ is shown on the x axis.
The accretion rate is given in dimensionless units.\footnote{In
these units the rate of the donor's mass loss is equal to 0.355
for the solution with a disc and 0.535 for the solution with a
cone shock.} From the moment when the matter with increased
velocity reaches the vicinity of the accreting component, the
accretion rate begins to grow and reaches its maximum after
approximately $0.06P_{orb}$ ($\sim$47 days). The maximum value of
the accretion rate is 2-2.2 times as compared to the state with
disc accretion.

As shown above, for accretor of 0.6$M_\odot$ mass, the ratio of
the upper limit of the steady burning range to its lower limit
$\mdot_{steady,max}/\mdot_{steady,min}\approx 2.2$. For a white
dwarf of 0.55$M_\odot$ mass (the value given in [\ref{MK92}]) an
increase of 10\% only should be enough to exceed the upper limit.
So, the approximately twofold accretion rate growth obtained in
our calculations is large enough to transfer the system from the
quiescent to active state.

The analysis of the data presented in Figures 4a,b shows that the
time of full disc destruction is $\sim180$ days. The time during which
the accretion rate exceeds the upper limit of the steady burning
range should correspond to the time of outburst development. If we
suppose that the exceedance of the limit of the steady burning
interval occurs when the accretion rate increases by 1.5 times,
the time during which accretion rate is above this value will be
approximately 100 days and $\approx2\times10^{-8}~M_\odot$
will be accreted during this time.

\bigskip

\section{Conclusions}

\bigskip

Outbursts in classical symbiotic stars occur when the accretion
rate becomes greater than the upper limit of the steady burning
range. For transition from quiescent to active state it is
necessary to provide an increase in the accretion rate on a rather
long time interval corresponding to the characteristic time of the
outburst development  ($\sim$100 days). Earlier, in
[\ref{Bisikalo2002}], a mechanism providing the required accretion
rate increase in the system has been proposed on the basis of 2D
calculations. According to this mechanism even minor change of the
donor's wind velocity is enough for the accretion regime to
change. During the transition from disc to wind accretion the
accretion disc is destroyed and the wind with increased velocity
makes the matter of the disc fall onto the accretor's surface. The
analysis of 2D calculations has shown that during this process the
accretion rate growth is enough for system to overstep the limits
of steady burning, therefore, leading to the development of an
outburst.

In order to answer if this mechanism can work in observed
astrophysical objects, the gasdynamic modelling of the flow
structure in the classical symbiotic system Z~And using more
realistic 3D model has been carried out. The results of our
calculations allowed us to draw following conclusions:

\begin{enumerate}

\item It was found that for the donor wind velocity 20 km/s
and $\Gamma=0.85 - 0.95$ the accretion disc forms in the system.
The accretion rate for the solution with disc is
$\sim22.5-25\%$ of the matter lost by the donor.

\item If the wind velocity equals 30 km/s the accretion disc disappears
and a cone shock forms. The accretion rate for this case is
$\sim11-13\%$ of the matter that left the donor.

\item These two solutions show that even minor change in the
donor's wind velocity (within the limits of $\pm5$km/s from the
observed value 25 km/s [\ref{FC88}]) leads to the change in flow
structure and to the accretion regime change, namely, to the
transition from the disc accretion to the wind accretion.

\item In accordance with the proposed mechanism, the
solution with the accretion disc should correspond to a quiescent
state of the system. According to our results, the accretion rate
in this case is $\sim22.5-25\%$. So, for Z~And, where the mass
loss rate is estimated to be $\sim2\cdot10^{-7}M_\odot$/yr, the
accretion rate will be $\sim4.5-5\cdot10^{-8}M_\odot$/yr,
in good agreement with the steady burning interval for
the most probable white dwarf masses.

\item In the solution where the wind velocity increases from
20 to 30 km/s the accretion disc is destroyed and  the
matter it contained falls onto the accretor's surface.

\item The detailed study of the accretion regime
change after the wind velocity increase has shown that the process
of disc destruction is followed by an accretion rate jump. The
value of maximum accretion rate is $\sim$2-2.2 times as much as
the initial value. As it has been mentioned above, the steady
burning range is rather narrow and if the accretor's mass equals
0.6$M_\odot$ the increase of the accretion rate by 2.2 times is
unambiguously enough for overstepping the upper limit of this
range.  Results of our calculations give an accretion rate
increase large enough to put the system outside the steady burning
range and to move it up into the active state.

\item The time during which the accretion rate is above the
steady burning range should correspond to the outburst development
time, that is approximately 100 days for Z~And. According to our
results the full time of disc destruction is approximately
$\sim180$ days. If we assume that the system leaves the steady
burning range after the accretion rate increases in 1.5 times, the
corresponding time will be $\sim100$ days, in good
agreement with observations.

\item The typical amplitude of Z~And outburst is  $\sim3^m$.
For a white dwarf of 1.2$M_\odot$ mass such brightness increase is
provided by an envelope mass $\sim 4\times10^{-7}$$M_\odot$
[\ref{PR80}]. In our calculations the envelope mass is
smaller, $\sim 10^{-8}$$M_\odot$.\footnote{It should
be noted that in the solution obtained the mass of the accreted
envelope depends on the model parameters (wind velocity, disc
temperature and density etc.) and is, therefore, not a fixed
value.} It is obvious, that for correct comparison of envelope
masses the calculations of thermonuclear burning on the surface of
a white dwarf with the mass equal to that of the accretor in Z~And
are required. Moreover it is necessary to take into account the
asymmetry of accretion. Unfortunately, we do not know any works
where such estimations were made, so, the question on the envelope
mass remains open. Given the accretion rate corresponding
to the upper limit of the steady burning range and given the typical
time of the outburst of $\sim 1/3$ year, the envelope mass should
not exceed significantly  $\sim 10^{-8}$$M_\odot$,
in good agreement with our results.

  \end{enumerate}

The analysis of results presented here allows us to draw the main
conclusion that the transition from quiescent to active state in
symbiotic stars can be concerned with the accretion regime change
(transition from the disc accretion to wind accretion) as the
result of insignificant variations of donor's wind velocity.

\medskip

This work was supported by the "21-st Century COE Program of
Origin and Evolution of Planetary Systems" of the Ministry of
Education, Culture, Sports, Science and Technology (MEXT) of
Japan. In particular, T.M., D.V.B. and H.B. acknowledge its
financial support. T.M. was supported by the grant in aid for
scientific research of the Japan Society of Promotion of Science
(13640241). Calculations were performed on the NEC SX-6
super-computer at the ISAS/JAXA Japan. The work was also
supported by Russian Foundation for Basic Research  (projects
03-02-16622, 03-01-00311, 05-02-16123, 05-02-17070, 05-02-17874),
by Science Schools Support Program (project N 162.2003.2), by
Presidium RAS Programs ``Mathematical modelling and intellectual
systems'', ``Nonstationary phenomena in astronomy''.

\section*{References}

\begin{enumerate}

\item
\label{Boyar67} A.~A.~Boyarchuk, Izv. CRAO {\bf 38}, 155 (1967).

\item
\label{FC88} T.~Fernandez-Castro, A.~Cassatella, A.~Gimenez, {\it
et al.}, Astrophys. J. {\bf 324}, 1016 (1988).

\item
\label{PR80} B.~Paczy\'nski and B.~Rudak, Astron. \& Astrophys.
{\bf 82}, 349 (1980).

\item
\label{MK92} J.~Miko\l ajewska and S.~J.~Kenyon, Monthly Notices
Roy. Astron. Soc. {\bf 256}, 177 (1992).

\item
\label{Boyar69} A.~A.~Boyarchuk, Izv. CRAO {\bf 39}, 124 (1969).

\item
\label{Kenyon86} S.~J.~Kenyon, The Symbiotic Stars, Cambridge
University Press, Cambridge (1986).

\item
\label{Baratta} G.~B.~Baratta, A.~Cassatella, and R.~Viotti,
Astrophys. J. {\bf 187}, 651 (1974).

\item
\label{Mammano} A.~Mammano and F.~Ciatti, Astron. \& Astrophys.
{\bf 39}, 405 (1975).

\item
\label{Ciatti} F.~Ciatti, A.~Mammano, and A.~Vittone, Astron. \&
Astrophys. {\bf 68}, 251 (1978).

\item
\label{Puetter}  R.~C.~Puetter, R.~W.~Russell, B.~T.~Soifer, and
S.~P.~Willner, Astrophys. J. {\bf 223}, L93 (1978).

\item
\label{Allen} D.~A.~Allen, Monthly Notices Royal Astron. Soc. {\bf
192}, 521 (1980).

\item
\label{Kenyon83} S.~J.~Kenyon and J.~W.~Truran, Astrophys. J. {\bf
273}, 280 (1983).

\item
\label{IbenTu96} I.~Iben,~Jr. and A.~V.~Tutukov, Astrophys. J.
Suppl. Ser. {\bf 105}, 145 (1996).

\item
\label{TU} A.~V.~Tutukov and L.~R.~Yungelson, Astrofiz. {\bf 12},
521 (1976).

\item
\label{Zytkow} B.~Paczy\'nski and A.~\.Zytkow, Astrophys. J. {\bf
222}, 604 (1978).

\item
\label{Iben} I.~Iben,~Jr., Astrophys. J., {\bf 259}, 244 (1982).

\item
\label{IbenTu89} I.~Iben,~Jr. and A.~V.~Tutukov, Astrophys. J.
{\bf 342},430 (1989).

\item
\label{Sion} E.~M.~Sion, M.~J.~Acierno, and S.~Tomczyk, Astrophys.
J. {\bf 230}, 832 (1979).

\item
\label{Fuji} M.~Y.~Fujimoto, Astrophys. J. {\bf 257}, 767 (1982).

\item
\label{Sienk} R.~Sienkiewicz, W.~Dziembowski, 1977, IAU Coll. {\bf
42}, 327 (1977).

\item
\label{Bisikalo2002} D.~V.~Bisikalo, A.~A.~Boyarchuk,
E.~Yu.~Kilpio, and O.~A.~Kuznetsov, Astron. Reports {\bf 46}, 1022
(2002).

\item
\label{Pacz} B.~Paczy\'nski, Acta Astron. {\bf 20}, 47 (1970).

\item
\label{Uus} U.~\"U\"us, Nauchn. Inf. {\bf 17}, 32 (1970).

\item
\label{GS} G.~S.~Bisnovatyi-Kogan, Ya.~M.~Kazhdan, A.~A.~Klypin,
A.~E.~Lutskii, Sov. Astron. {\bf 233}, 201, (1979).

\item
\label{Dima94} D.~V.~Bisikalo, A.~A.~Boyarchuk, O.~A.~Kuznetsov
{\it et al.},  Astron. Reports, {\bf 38}, 494 (1994)

\item
\label{Dima96} D.~V.~Bisikalo, A.~A.~Boyarchuk, O.~A.~Kuznetsov,
and V.M.Chechetkin,  Astron. Reports, {\bf 40}, 662 (1996)

\item
\label{gamma1} K.~Sawada, T.~Matsuda, and Hachisu, Monthly Notices
Roy. Astron. Soc. {\bf 219}, 75 (1986).

\item
\label{gamma2} D.~Molteni, G.~Belvedere, and G.~Lanzafame, Monthly
Notices Roy. Astron. Soc. {\bf 249}, 748 (1991).

\item
\label{gamma3} D.~V.~Bisikalo, A.~A.~Boyarchuk, O.~A.~Kuznetsov
{\it et al.},  Astron. Reports, {\bf 39}, 325 (1995)

\item
\label{Nagae} T.~Nagae, K.~Oka, T.~Matsuda {\it et al.}, Astron.
\& Astrophys. {\bf 419}, 335 (2004).

\item
\label{Makita} M.~Makita, K.~Miyawaki, T.~Matsuda, Monthly Notices
Roy. Astron. Soc. {\bf 316}, 906 (2000).

\item
\label{Fujiwara} H.~Fujiwara, M.~Makita, T.~Matsuda, Progress of
Theoretical Physics {\bf 106}, 729 (2001).

\item
\label{MM98} N.~Mastrodemos, M.~Morris, Astrophys. J. {\bf 497},
303 (1998).

\item
\label{WF99} R.~Walder, D.~Folini, Thermal and Ionization Aspects
of Flows from Hot Stars, Edited by Henny Lamers and Arved
Sapar,ASP Conference Series, {\bf 204}, 331 (2000)

\item
\label{Dumm} T.~Dumm, D.~Folini, H.~Nussbaumer {\it et al.},
Astron. \& Astrophys. {\bf 354}, 1014 (2000).

\end{enumerate}

\clearpage
\begin{center}

FIGURE CAPTIONS

\medskip
 for the paper by Bisikalo et al "3D Gasdynamic Modelling \dots"
\end{center}

\vspace{0.5cm}

\medskip\noindent {\bf Fig.~1.}
Density contours and velocity vectors in the  equatorial plane of
the system for the case $V_w=20$ km/s and $\Gamma$=0.94. The empty
circle centered at (-1,0) corresponds to the donor (radius of the
circle equals donor's radius), the diamond at (0,0) point marks
the accretor.

\medskip\noindent {\bf Fig.~2.}
The same as in the Fig.~\ref{fig_1} in the vicinity of the
accretor.

\medskip\noindent {\bf Fig.~3.}
Density contours and velocity vectors in the equatorial plane of
the system for the case $V_w=30$ km/s and $\Gamma$=0.94. All
designations are the same as in Fig.~\ref{fig_1}.

\medskip\noindent {\bf Fig.~4.}
The same as in Fig.~\ref{fig_3} in the vicinity of the accretor.

\medskip\noindent {\bf Fig.~5.}
Density contours and velocity vectors in the area near the
accretor $\sim$180 days after the wind velocity increase from 20
to 30 km/s.

\medskip\noindent {\bf Fig.~6.}
Accretion rate change for the solution where the wind velocity is
increased from 20 to 30 km/s. The vertical line marks the moment
of the wind velocity change.

\medskip\noindent {\bf Fig.~7.}
The same as in Fig.~\ref{fig_6} for short time interval beginning
from the moment of wind velocity increase from 20 to 30 km/s.


\clearpage

\renewcommand{\thefigure}{1}
\begin{figure} [!ht]
\centerline{\hbox{\epsfig{figure=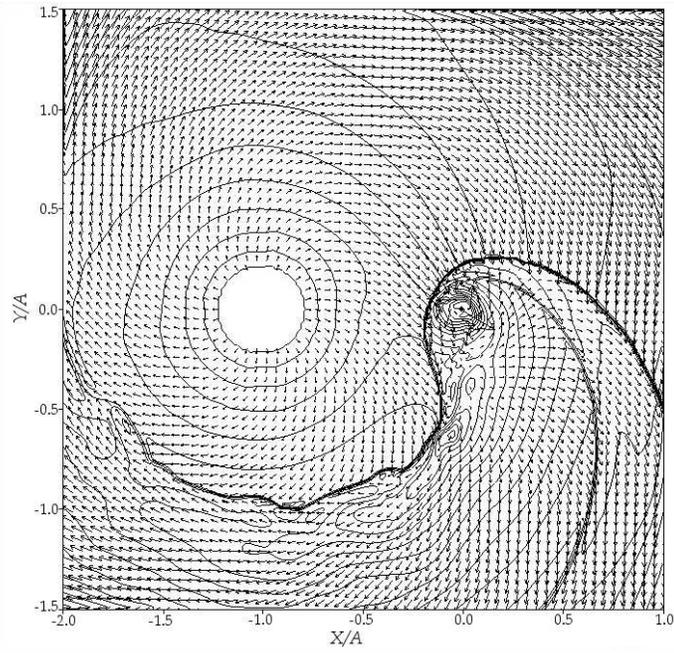,width=10cm}}}
\vspace{5mm} \caption{\small For the paper by Mitsumoto et al. "3D
Gasdynamic Modelling \dots"} \label{fig_1}
\end{figure}

\renewcommand{\thefigure}{2}
\begin{figure} [!ht]
\centerline{\hbox{\epsfig{figure=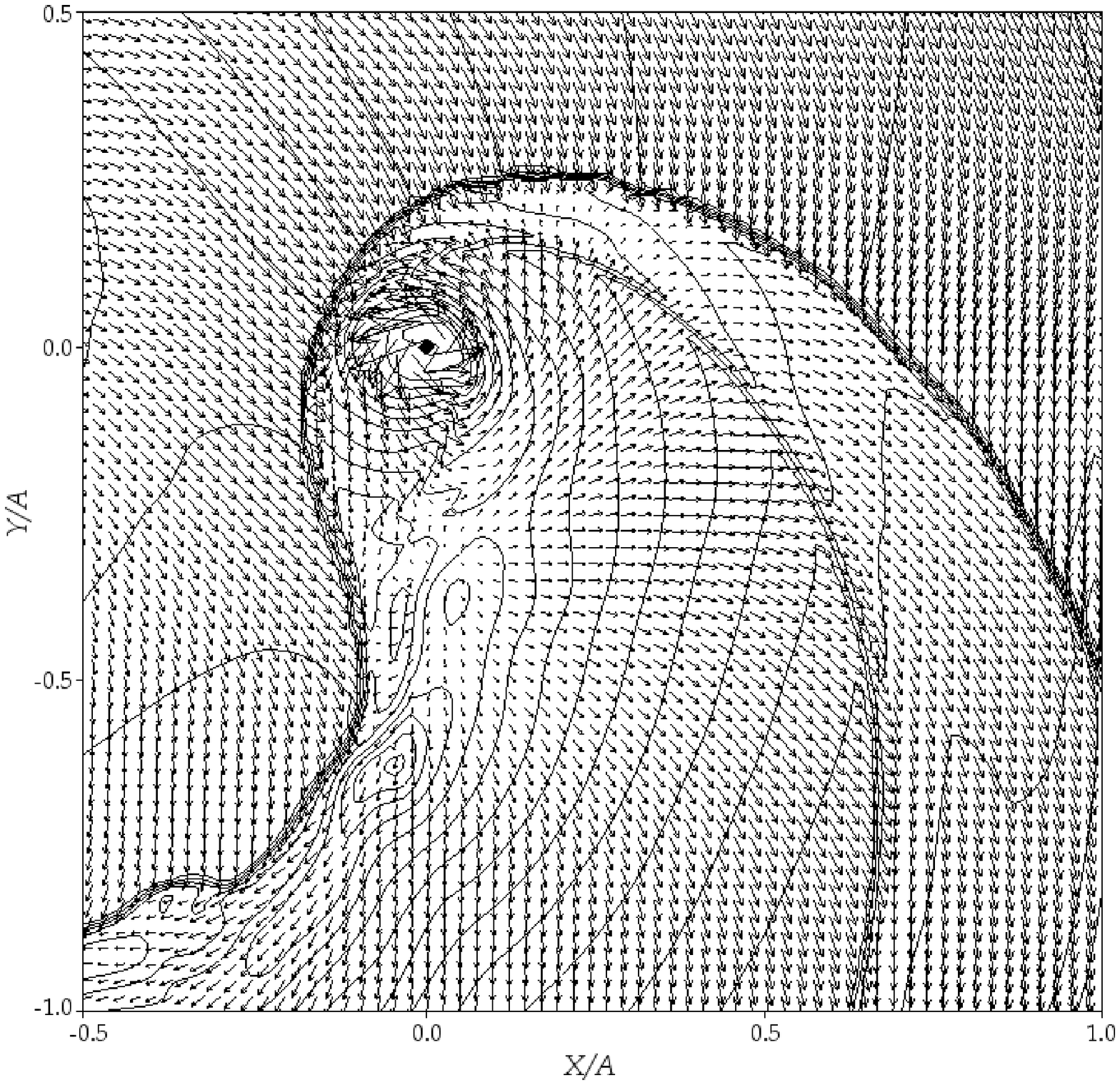,width=10cm}}}
\vspace{5mm} \caption{\small For the paper by Mitsumoto et al. "3D
Gasdynamic Modelling \dots"}  \label{fig_2}
\end{figure}

\renewcommand{\thefigure}{3}
\begin{figure} [!ht]
\centerline{\hbox{\epsfig{figure=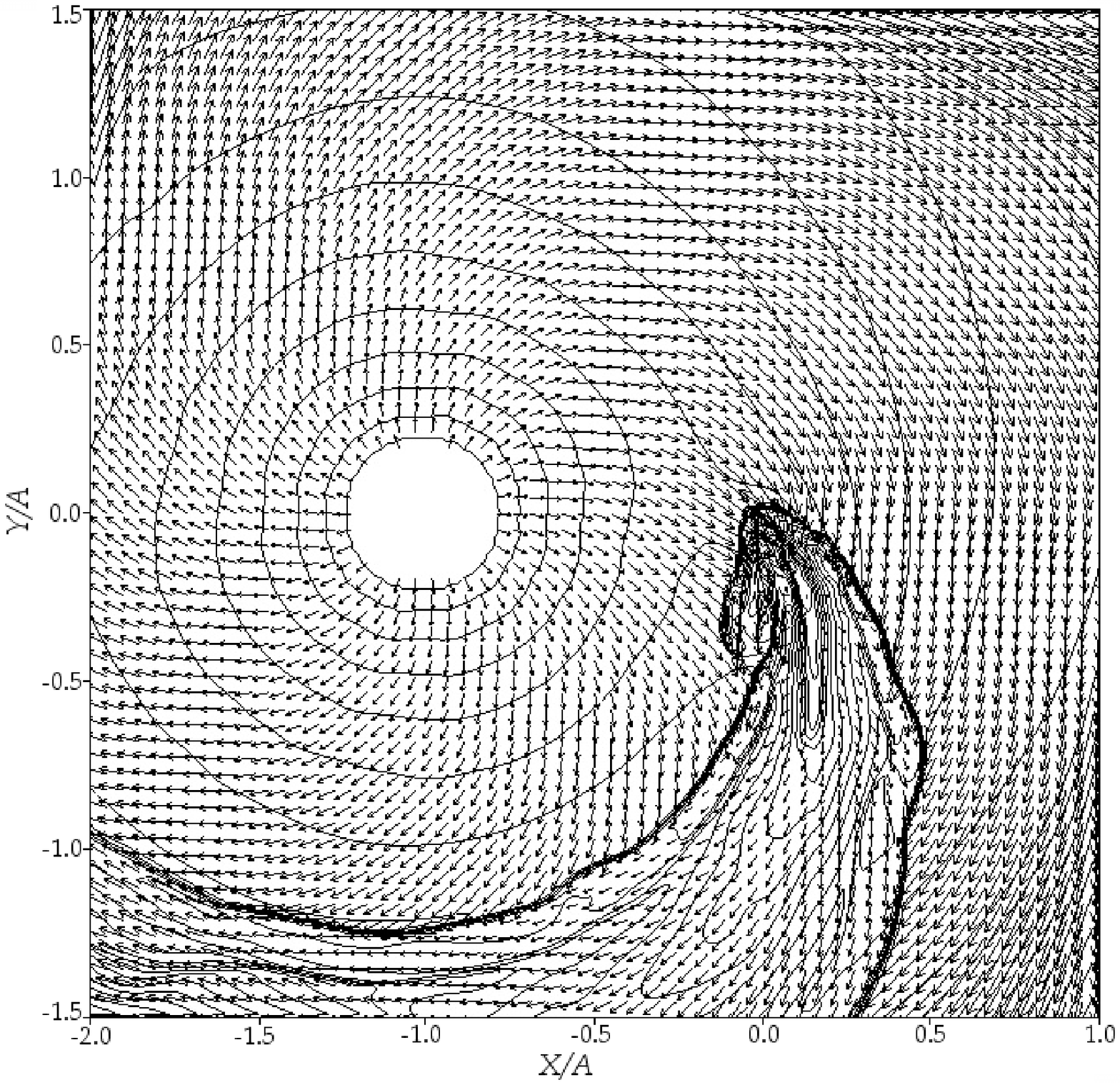,width=10cm}}}
\vspace{5mm} \caption{\small For the paper by Mitsumoto et al. "3D
Gasdynamic Modelling \dots"} \label{fig_3}
\end{figure}

\renewcommand{\thefigure}{4}
\begin{figure} [!ht]
\centerline{\hbox{\epsfig{figure=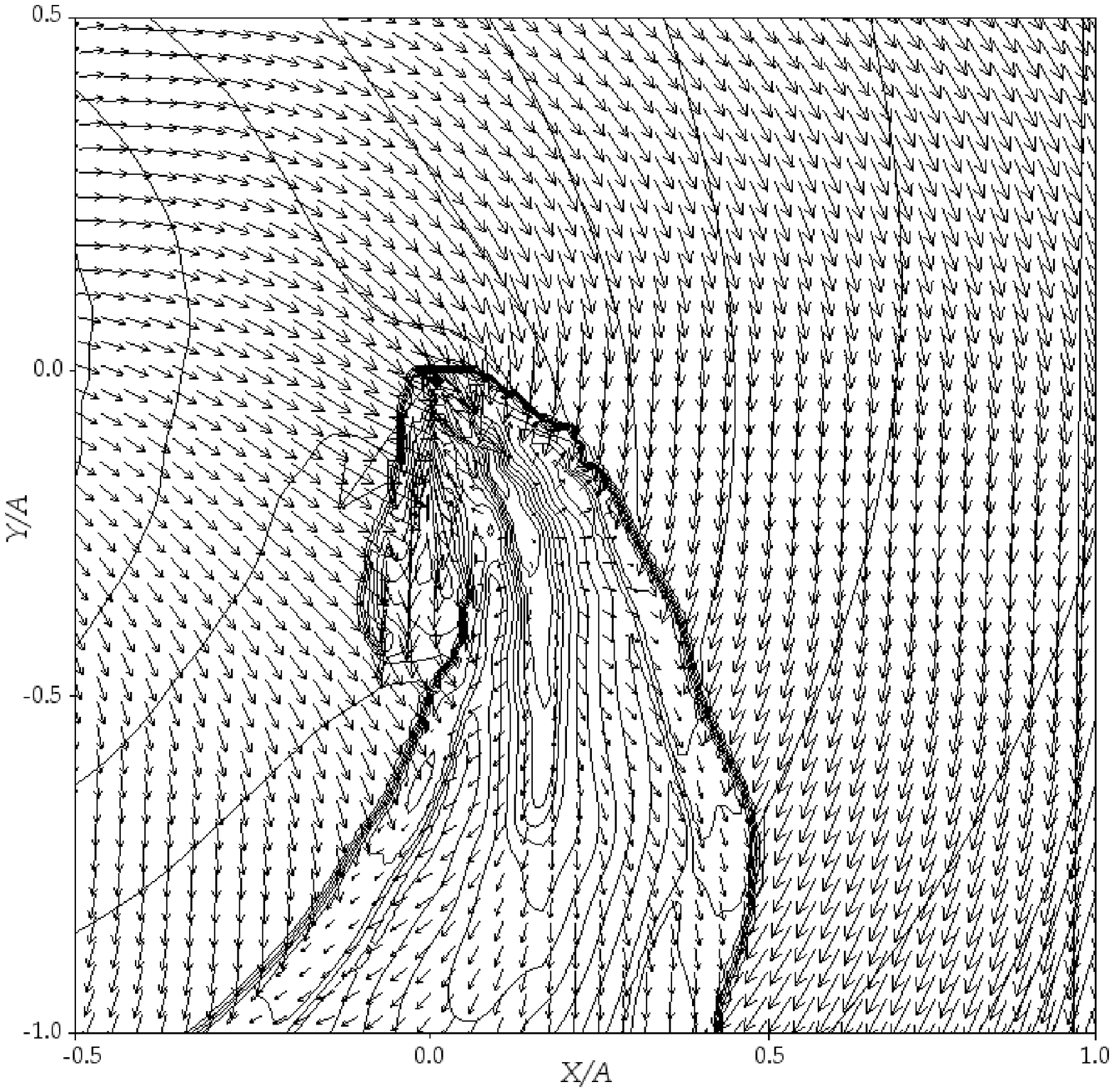,width=10cm}}}
\vspace{5mm} \caption{\small For the paper by Mitsumoto et al. "3D
Gasdynamic Modelling \dots"}  \label{fig_4}
\end{figure}

\renewcommand{\thefigure}{5}
\begin{figure}[!ht]
\centerline{\hbox{\epsfig{figure=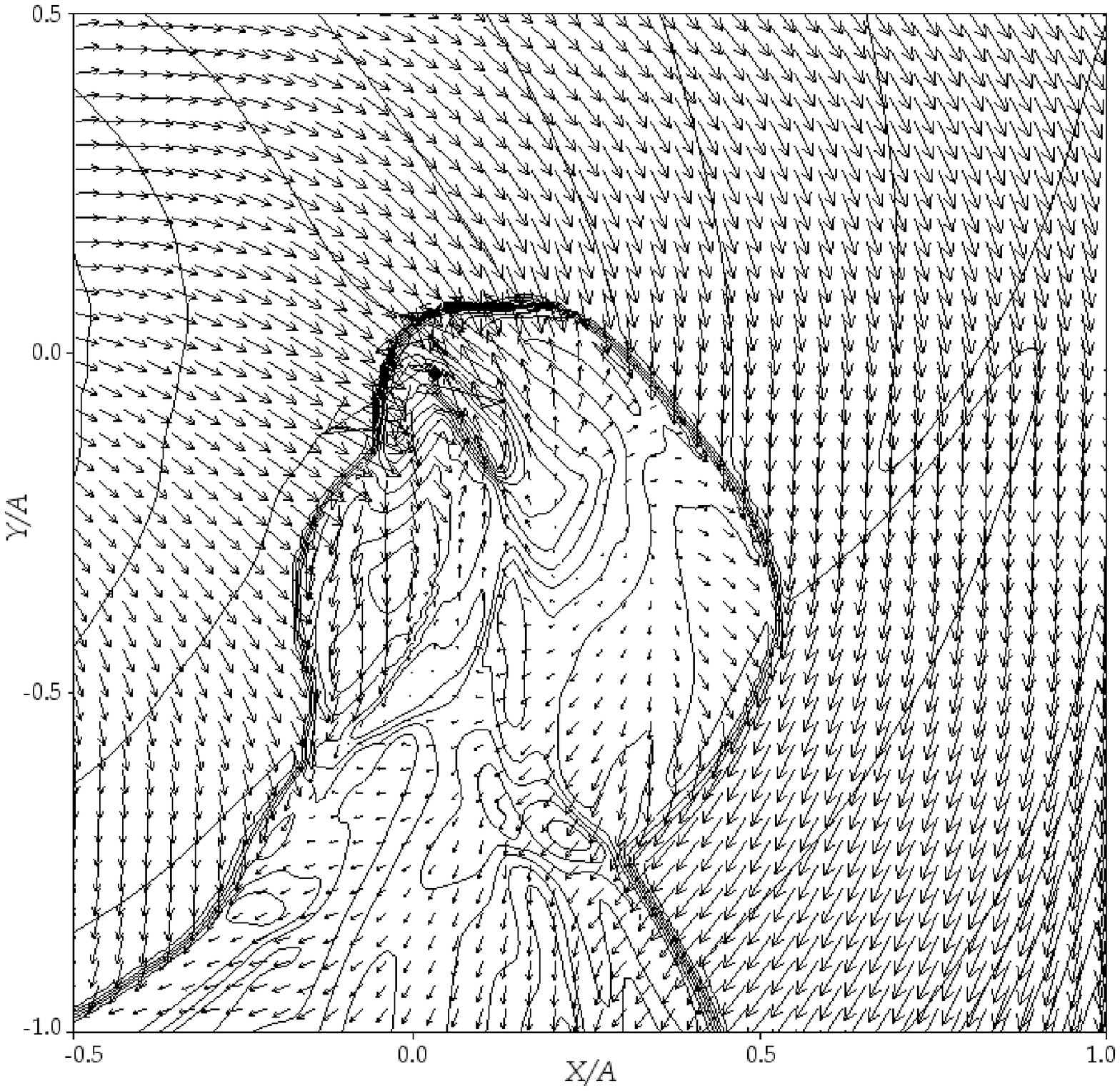,width=10cm}}}
\vspace{5mm} \caption{\small For the paper by Mitsumoto et al. "3D
Gasdynamic Modelling \dots"} \label{fig_5}
\end{figure}

\renewcommand{\thefigure}{6}
\begin{figure}[!ht]
\centerline{\hbox{\epsfig{figure=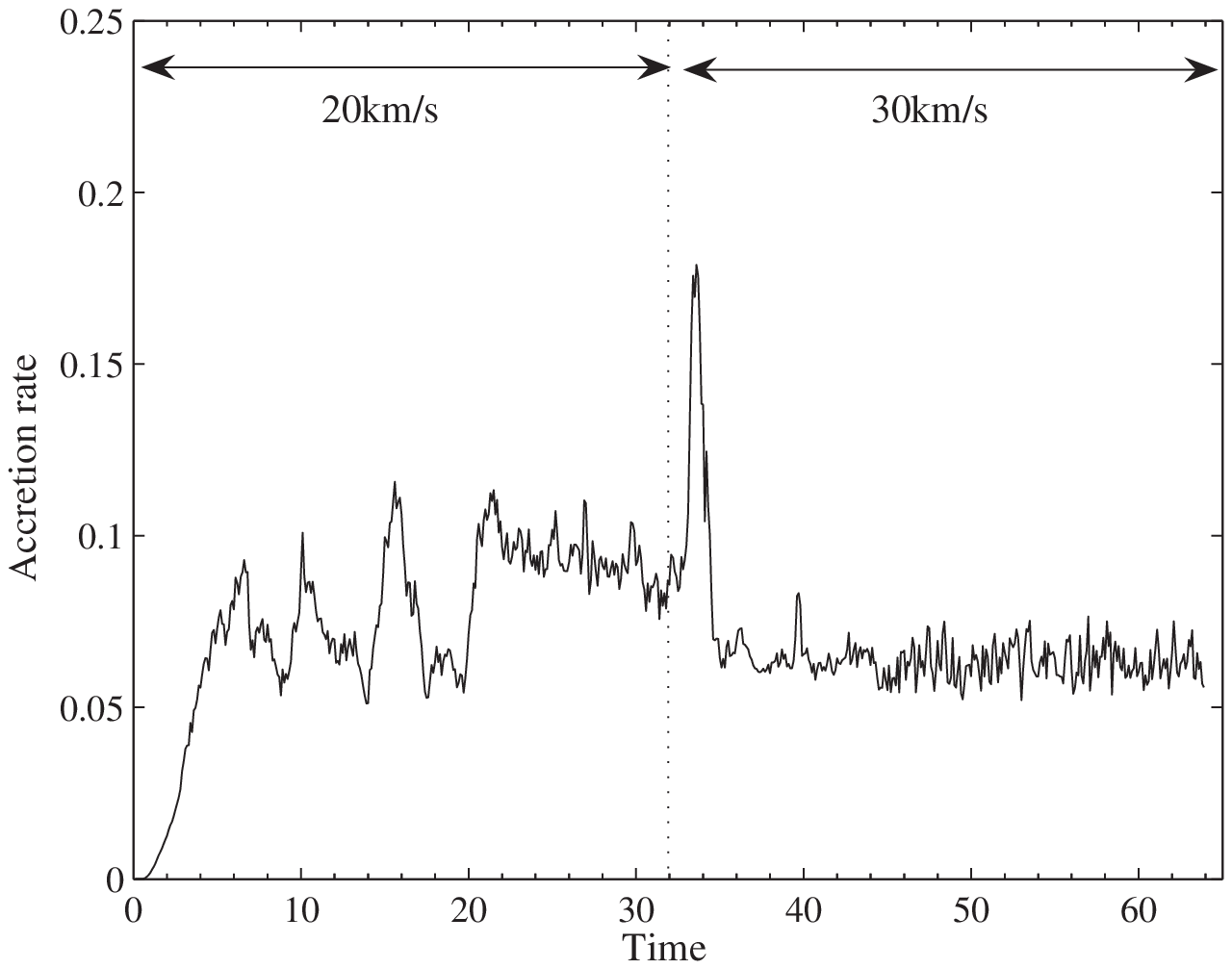,width=10cm}}}
\vspace{5mm} \caption{\small For the paper by Mitsumoto et al. "3D
Gasdynamic Modelling \dots"} \label{fig_6}
\end{figure}

\renewcommand{\thefigure}{7}
\begin{figure}[!ht]
\centerline{\hbox{\epsfig{figure=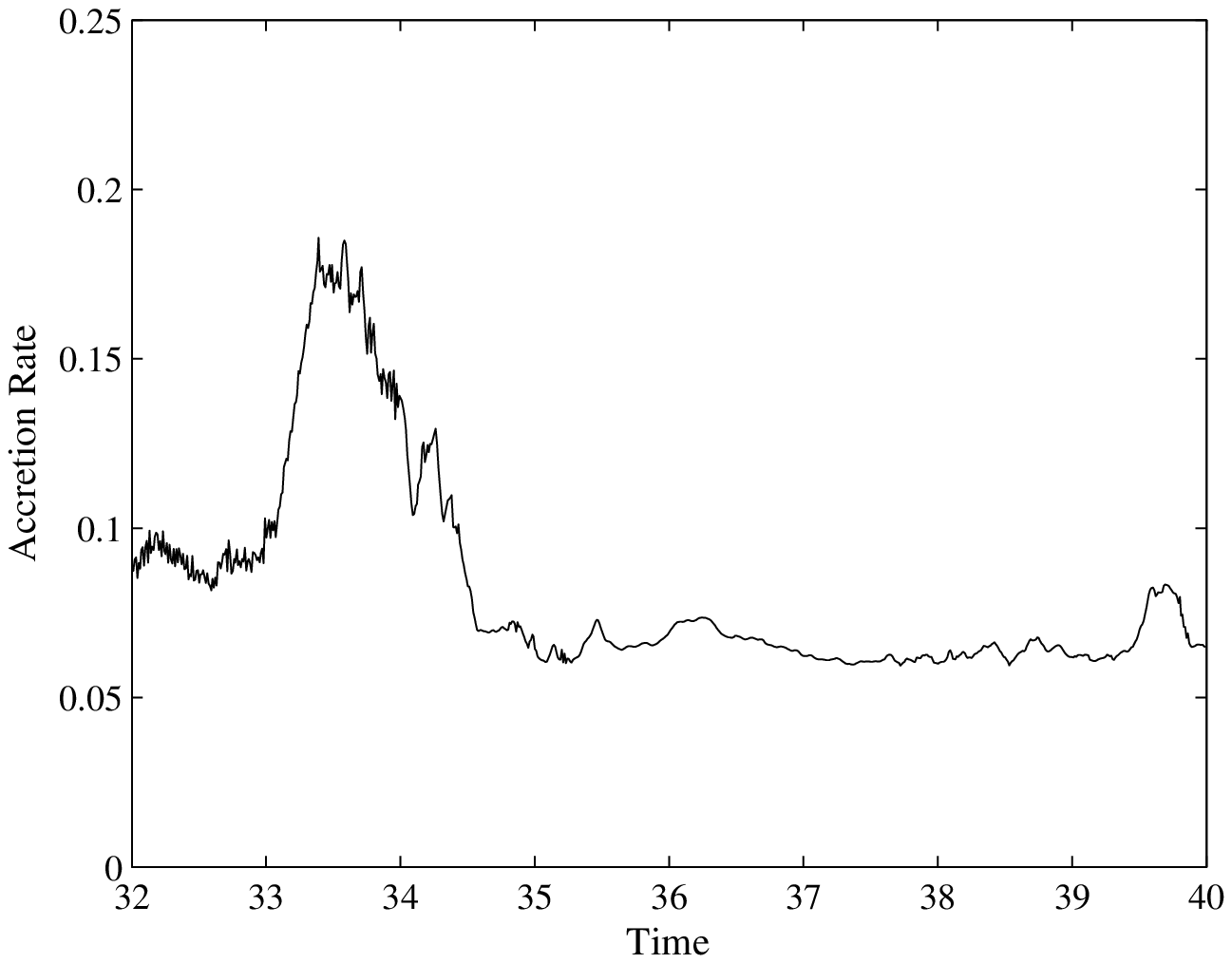,width=10cm}}}
\vspace{5mm} \caption{\small For the paper by Mitsumoto et al. "3D
Gasdynamic Modelling \dots"} \label{fig_7}
\end{figure}

\end{document}